# Enabling Student Innovation through Virtual Reality Development

Sherri Harms
Department of Computer Science and Information Technology
University of Nebraska at Kearney
Kearney, NE 68849
harmssk@unk.edu

## Abstract

It is clear, from the major press coverage that Virtual Reality (VR) development is garnering, that there is a huge amount of development interest in VR across multiple industries, including video streaming, gaming and simulated learning. Even though PC, web, and mobile are still the top platforms for software development, it is important for university computer science (CS) programs to expose students to VR as a development platform.

Additionally, it is important for CS students to learn how to learn about new technologies, since change is constant in the CS field. CS curriculum changes happen much slower than the pace of technology adoption. As new technologies are introduced, CS faculty and students often learn together, especially in smaller CS programs. This paper describes how student-led VR projects are used, across the CS curriculum, as basic CS concepts are covered. The student-led VR projects are engaging, and promote learning and creativity. Additionally, each student project inspires more students to try their hand at VR development as well.

Sherri Harms
Department of Computer Science and Information Technology
University of Nebraska at Kearney
Kearney, NE 68849
harmssk@unk.edu

# 1 Introduction

The Computer Science 2013 curriculum guidelines for undergraduate programs in computer science [1] state that "*all graduates of computer science programs should have been involved in at least one substantial project.*" and "*graduates should realize that the computing field advances at a rapid pace, and graduates must possess a solid foundation that allows and encourages them to maintain relevant skills as the field evolves.*" "*To develop this ability, students should be exposed to multiple programming languages, tools, paradigms, and technologies as well as the fundamental underlying principles throughout their education.*"

There are many ways computer science programs meet these goals and expose students to new technologies in the curriculum. Many educators have developed special topic courses that employ new technologies or use a project-driven approach to engage and motivate students. Pragmatically, most departments are under pressure to reduce course delivery costs, but techniques to motivate all students usually increase costs because they require the development of new material, running multiple level classes, and offering differentiated material [2].

Denning and McGettrick [3] argue that computer science should be "*recentered around a theme on innovating*". They state that "*innovation can be learned and a spirit of innovation can permeate throughout our courses, starting from the first year and building up expertise in innovation through to the final year.*"

One of the best ways to introduce creativity and innovation is through the use of project-based learning (PBL) [4]. Straub summarized: *PBL can provide ancillary benefits such as improved student creativity, self-image, and motivation. It also has been shown to have workforce preparation, job placement, academic program retention, and knowledge retention benefits* [4]. Student-led projects (SLPs) are a form of PBL which offer the greatest degree of flexibility for student participants. These types of projects may start in response to an external stimulus (such as a design competition or other program) or simply based on students' desire to learn more about or work with a topic.

At the University of Nebraska at Kearney (UNK), the Computer Science and Information Technology (CSIT) programs utilize SLPs in many ways, across the CS curriculum, embedded within several conventional courses and as part of the senior design experience course [5]. As a small CS program, using SLPs allows students to work with new technologies, without having a specialized elective course on a topic. Using SLPs has instilled a culture of innovation and creativity embraced by the CS student body, where advanced students assist newer students as they embark on their journey.

# 2 Virtual Reality (VR)

There is a huge amount of development interest in VR across multiple industries, including video streaming, gaming and simulated learning. Even though PC, web, and mobile are still the top platforms for software development, it is important for university computer science programs to expose students to VR as a development platform.

Using VR as a development platform requires students to use innovation, creativity, and problem solving skills. VR replicates an environment that simulates a physical presence in places in the real world or an imagined world, allowing the user to interact in that world. Virtual realities artificially create sensory experiences, which can include sight, touch, hearing, and smell.

VR replicates an environment that simulates a physical presence in the real world or an imagined world, allowing the user to interact in that world. Virtual realities artificially create sensory experiences, which can include sight, touch, hearing, and smell. The simulated environment can be similar to the real world in order to create a lifelike experience—for example, in simulations for pilot or combat training—or it can differ significantly from reality, such as in VR games. Most advances are being done in the entertainment industry, but many understand and realize the future and the importance of VR in many fields, especially in education, training and simulation, tourism, patient care, mental therapy, architecture and modeling [6].

VR was introduced in 1950 when Morton Heilig first described his idea for an "*experience theatre*," and it gained some media attention in the 1990s. Due to the expensive and time-intensive development cycle and poor graphics, VR fell into a hiatus until 2012, when Palmer Luckey introduced the Oculus Rift through a Kickstarter campaign [7]. Unexpectedly, the campaign raised more than $670,000 in its first 24 hours, and more than a million dollars within three days. Rift has received a great deal of industry and consumer attention for its high-quality resolution, low latency, and affordability [8].

In the three years since its introduction, Oculus has gone on to be acquired by Facebook for two billion dollars, forge partnerships with Microsoft and Samsung, widen its scope, and ultimately become the poster boy for VR [9]. Developer support for Oculus Rift and other virtual reality headsets is at an all-time high [10]. Oculus' momentum clearly demonstrates technology has reached a point, in terms of quality and price, where these devices can be made available to the public in an attractive format. And, more importantly, they demonstrate new possibilities that can spark our imagination [11].

The 2015 GDC survey [10] revealed that development of virtual reality (VR) titles has more than doubled among participating developers with 16 percent currently developing for VR. This is up from the 7 percent of developers were working on VR projects in last year's survey. Additionally, the looming launch of Oculus Rift, HTC Vive, and PlayStation VR has clearly begun to sway developers [10].

Developers are also pretty keen on VR being a long-term platform, with 75 percent stating it as a long-term sustainable business. Yet, GDC's survey shows that many developers don't see VR becoming commonplace until 2030 [9]. Last November, analysts at Sophic Capital predicted VR will reach a $7 billion market by 2018. Others estimate up to $62 billion by 2025 [11].

## 3 Use of VR Student-Led Projects in a CS program

Given the newness of the hardware and software tools used for VR, it is hard for faculty and students to develop an accurate project management plan for successful implementation, especially for a projects that contribute heavily to a semester course grade. The fear of the unknown can prohibit from students from even considering a VR project, especially if "easier,"

"safer" alternatives can be selected to meet the course requirements. Additionally, inexperience with VR tools can cause students who start a VR project to get overwhelmed and lose the confidence and drive to complete the project.

For cutting edge student-led projects, such as VR projects, the risk of project failure is high. For students to be successful, risks must be managed. The key risks are with estimating the time projected to complete tasks; estimating task complexity; identifying required tasks; and identifying knowledge, skill and experience gaps of students (and faculty) related to developing for a VR platform.

To mitigate the risks students are willing to take, students should be reassured that a failed project can still be a fully successful SLP. It is only by failing to try that the student's grade is a failure as well. Students need to be encouraged to try something fun, unusual and interesting – even if they only get partway done. Instructors should provide sample projects from previous students that illustrate "A" projects that were not fully functional when completed. Additionally, the complexity of the project should be a graded component. Thus, students who choose a more complex project will not be penalized when they do not get all of the project components implemented. Similar to Lewandowski et al. [12] and Romeike [13], this approach increases student willingness to tinker with something new, and establishes a culture of innovation.

Instructors who use SLPs in their courses need to explain that this is the time in a CS student's life to experiment and "play" as they learn. Additionally, milestones should be used throughout the semester to help the students stay on track, and to provide plenty of graded components of the project. These components help students demonstrate their communication and other professional skills, and provide opportunities for them to manage their project and to reflect on the experience [14].

For example, in the Artificial Intelligence (AI) class at UNK, as the semester project is explained, students must submit a brief proposal with basic ideas and resources they will be using. The instructor reviews these proposals to gage if the project is doable within the semester timeframe, and if the project meets the overall goal of the assignment. The proposal may need to be resubmitted numerous times before it is approved by the instructor.

The projects that students propose must have some core features that demonstrate an understanding and implementation of the course material while at the same time leaving room for individual creativity. Students are not told which technology to use. They may choose from many different technologies, VR being one of the more popular one. Mobile app and game development are other common choices for SLPs.

Students in the AI class give a background presentation to the class part-way through the semester. Students introduce the goals of their project, describe the foundational course material being implemented and describe the resources they are using in their project. Students give each other feedback and suggestions.

AI students also keep a weekly blog to report on their progress. These provide opportunities for the instructor and other classmates to provide feedback and suggestions as the semester progresses. According to Makiaho [15], o*ne of the most important reason that professional*

*software engineering project fails is poor reporting of the project's status. Because of the poor reporting, the management does not know the state of the project and thus does not execute the right actions.* Student software development projects can fail for same reason, thus blogging is a key component to the success of SLPs.

Blogging also provides students with a place for reflection and analysis. Students are required to submit at least one entry a week for the last five weeks of the semester, along with a final blog post during finals week. Blog posts should be well written and explain the project and the on-going progress. It should provide the background for the project and links or references for websites, articles, etc. that were used. Screen shots or pictures are used to enhance the blog postings. Students are expected to explain the algorithms and data structures used, along with the experiments conducted. They are also expected to explain lessons learned, discuss future enhancements that should be made, and future testing that should be performed. Thus, blogging becomes an important tool for developing student soft skills.

Straub [16] found that a key risk mitigation factor was "*ensuring that project knowledge is documented in a known and accessible location.*" Students use code repositories to manage code. Together with the blog and internal code documentation, the project knowledge is well documented and accessible.

At the end of the semester, AI students complete a code walk-through, a final project presentation, along with their final reflection blog post that also includes an executable download, a link to a video demonstrating the project, or some way for the viewer to see the project in action. (Rubrics for these outcomes are available upon request.)

The final project presentations are given in a professional conference style. Business representatives, alumni, and the entire CS student body are invited to attend. This provides an opportunity for students to disseminating results, gain recognition, and develop their oral communication skills.

## 4 Sample Virtual Reality Student-Led Projects

Starting in 2014, the UNK CSIT department adopted VR resources and lab space for student projects. Using VR lab resources, CSIT students quickly became innovators in the VR world. Each new version of VR hardware and/or software and each new semester, brought opportunities for students to try new things and express their creativity. Additionally, each student project inspires more students to try their hand at VR development.

Through this entire process, the faculty have fostered student innovation, but have not been the information providers – students have relied on online resources and each other for information. The faculty have provided students with opportunities to learn from authentic experiences, and allowed students to explore what is relevant to them.

The first CSIT VR project was a snowboarding simulation completed in the spring of 2014 as Software Engineering SLP. It used the first generation Oculus Rift VR display for head tracking and display along with the Wii Balance Board [17], to control the in-game snowboard. It was developed with the Unity Engine [18], and used Blender [19] for 3D modeling. An image of a student using the snowboarding simulation is shown in Figure 1 on the left, and a video demonstration is available on YouTube [20].

In the spring of 2015, two students created VR projects in the Artificial Intelligence (AI) course, using the Oculus Rift Development Kit (DK) 2. The first was an "robots versus humans" VR first person shooter game, [21] for the Oculus Rift KD2. A screen shot of this game is shown in Figure 1 on the right. It used the Unity Game Development Engine, along with TF3DM [22] and Turbo Squid [23] for 3D modeling. The most challenging aspect of the development process was making the robot have enough intelligence to target the human controlled player, but not enough intelligence that the human controlled character doesn't have a chance.

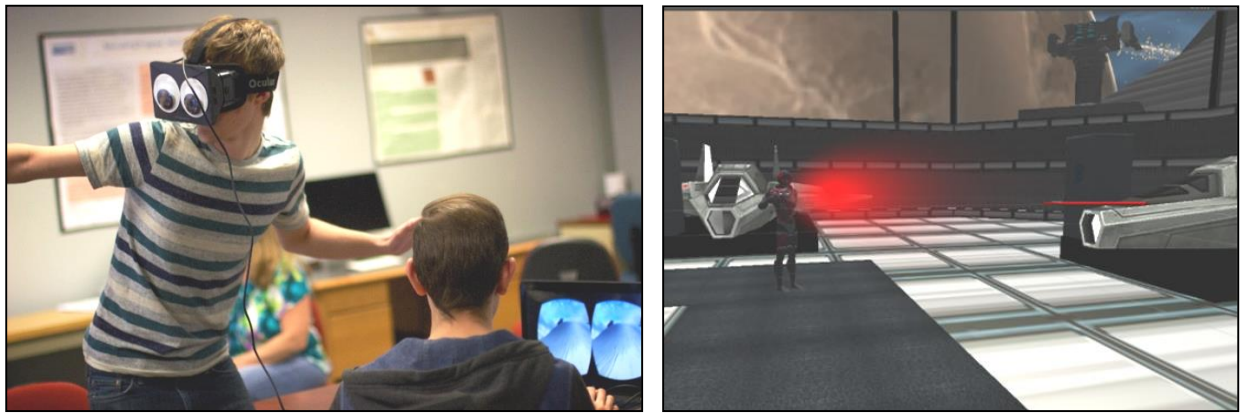

Figure 1: Sample Virtual Reality Student Led Projects

The second VR SLP created in the AI class, was a VR motorcycle racing game. The game involves a player surviving on a motorcycle while avoiding walls and an AI motorcyclist. Players wear the Oculus Rift DK2 headset to look and feel like they are actually riding a motorcycle. A video demonstration and explanation of the game is available online [24].

The fall 2015 semester brought about more student-led VR projects. The student who created the motorcycle racing game in AI, set out to improve it in the Computer Graphics course. Unfortunately, innovating in the VR world has its challenges. As the game was ported to new Unity 5 engine, the motorcycle which worked perfectly in the previous version, now moved straight up instead of straight forward. This problem was quickly fixed. The ability for the human player to jump over walls was added and other adjustments to the environment were made. Allowing a SLP to continue multiple semesters encourages students to start a cutting edge VR project, as they know it is worth the initial effort and time investment, and it allows them to get deeper into the technology and practice the process of iterative development.

In fall 2015, a team of two students created a VR project in their Senior Capstone course. This was a VR 3D boxing game for the Oculus Rift. It was built using Unity with Leap Motion, using Blender for the 3D modeling. The Leap Motion allows the VR system to "see" the player's hands. Unfortunately, it only recognizes a hand when all five fingers were visible. This made it difficult to see the player's fist, as needed for their boxing game. Their game turned into more of a "slapping" game. One of the students is shown in Figure 2 presenting the game at the CSIT Fall 2015 Project Demonstration Day [5].

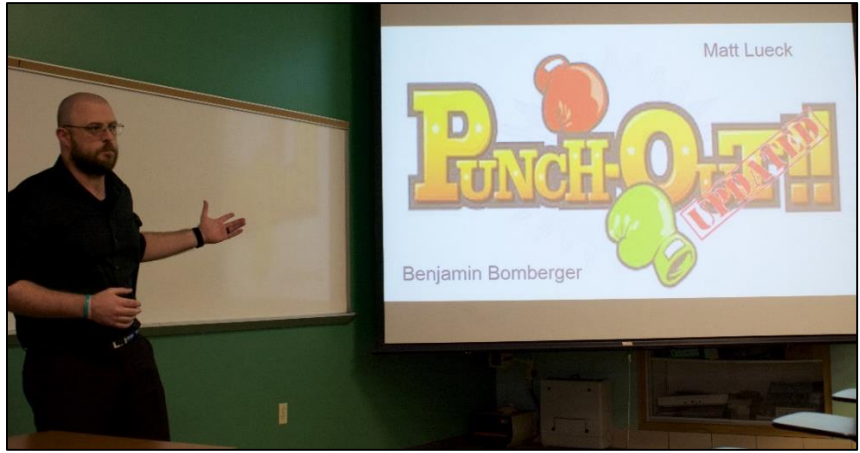

Figure 2: Sample Student Presentation on Virtual Reality Student Led Project

## 5 Results & Discussion

Using SLPs, including VR projects, within the context of a course, students are empowered with the opportunity to complete their own projects. Additionally, when lab time is given to work on these projects, the instructor is available to answer questions as they arise. This also helps the instructor make sure that the projects are on track to be completed. Students also tend to help one another as they learn something new. However, there is a limit to the number of SLPs that an instructor can manage. At UNK, it was found that using SLPs with a class of more than 30 students was hard to manage. Additionally, SLPs often require unique hardware/software purchases, which can be costly.

Romeike [13] found, "it was most rewarding for students to strive for good working software, based on self-chosen and meaningful tasks, even if the resulting products did not have any outside value." We found similar results.

After completing SLPs, students in the AI class commented, *"Allowing students to choose their projects allowed passion to develop which is extremely useful." "I learn the best by doing things. I'm glad this class implemented major projects instead of tests to reinforce knowledge." "I like the inclusion of the final project."* Software Engineering students commented, *"Being able to pick a project that is interesting to us was great." "I think actually doing a software engineering project ourselves was a good idea."* Senior Capstone students commented, *"I loved getting to pick a project of interest to me and having a semester to work on it." "Picking a project is a great choice for students. Learned a lot." "Enjoyed having the freedom to choose our own projects." "Cool Beans!  This was cool/Fun!"*

There are many advantages and benefits to supporting SLPs. Similar to undergraduate research, academic departments benefit from the reflected glory of successful projects which can be reported on web sites and in news articles. Student successes in SLPs help encourage and motivate other students as well.

For example, the snowboarding VR project was presented to local businesses and CSIT students at the CSIT Spring 2014 Project Demonstration Day; and at outreach events for middle and high school students at the 2014 Nebraska Broadband Conference and the 2015 First Lego League Robotics Competition in Kearney, and the motorcycle racing game was presented at an IT Expo [25] for high school students in March 2016, as shown in Figure 3 on the right.

The student who created the motorcycle racing game, Sam Middleton, is shown in Figure 3 on the left pitching his idea for a VR game development company. His idea won the 5th annual Central Nebraska business idea contest in October 2015 [26].

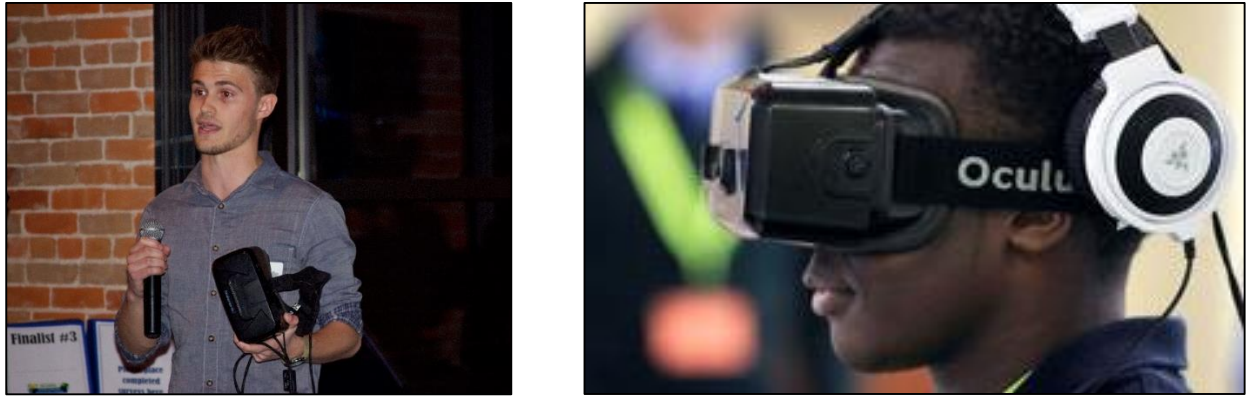

Figure 3: a.)Student Pitching VR Game Development as a Business
b.) High School Student Playing UNK CS Student's VR game at itx16.gips.org

"*With the rising market in virtual reality gaming, such a business would thrive in the coming years. And unlike the physical products of most businesses, games could be mass-produced cheaply through digital online marketplaces*," said Middleton [26].

Middleton continued, "*As these virtual reality devices are created by Microsoft, Sony, Samsung and Google, there will be a high demand for software compatible with these devices*" [26].

Notice that he is the same student trying out the VR project three semesters earlier, on the left in Figure 1. This illustrates how the students who complete VR SLPs inspire and teach the younger students to also try VR development.

## 6 Conclusion

VR will be a major contributor to new technologies and innovations for years to come. The UNK CSIT Department plans to stay on the leading edge of this field, proving the tools and resources needed to inspire student creativity and innovation so they can be successful in their VR classroom projects – enabling them to be successful entrepreneurs in VR.

Empowering students through the use of SLPs across the curriculum embodies the goals of the CS2013 curricula guideline. It is a holistic approach that goes beyond exposing students to technological facts. It helps students develop their verbal and written communication, time management, and problem solving skills. It helps students develop risk tolerance, innovation, and creativity.